\documentclass[aps,pra,onecolumn,groupedaddress,asmsymb,notitlepage,longbibliography]{revtex4-1}
\usepackage{mathtools,amsmath,amssymb}
\usepackage{graphicx}
\usepackage{color}
\usepackage[usenames,dvipsnames]{xcolor}

\newcommand{\bs}{\boldsymbol}
\newcommand{\bu}{{\bs u}}
\newcommand{\bbf}{{\bs f}}
\newcommand{\bbF}{{\bs F}}
\newcommand{\br}{{\bs r}}

\newcommand{\bx}{{\bs x} }

\newcommand{\bX}{{\bs X} }
\newcommand{\bw}{{\bs w} }
\newcommand{\bv}{{\bs v} }
\newcommand{\oell}{{\overline{\ell}}}
\newcommand{\tell}{{\tilde \ell}}

\newcommand{\otX}{{\widetilde{\overline{\bX}}}}
\newcommand{\otu}{{\widetilde{\overline{\bu}}}}
\newcommand{\oG}{{\overline{G}}}

\newcommand{\smallfilter}[1]{{\overline{#1}}}
\newcommand{\bigfilter}[1]{{\widetilde{#1}}}

\definecolor{jhublue}{RGB}{0,105,175}

\def\undertilde#1{\mathord{\vtop{\ialign{##\crcr
$\hfil\displaystyle{#1}\hfil$\crcr\noalign{\kern1.5pt\nointerlineskip}
$\hfil\tilde{}\hfil$\crcr\noalign{\kern1.5pt}}}}}

\begin{document}

\title{On the large-scale sweeping of small-scale eddies in turbulence -- A filtering approach}

	\author{Theodore D. Drivas}
	\affiliation{Department of Applied Mathematics \& Statistics, The Johns Hopkins University, Baltimore, MD 21218, USA }

	\author{Perry L. Johnson}
	\affiliation{Department of Mechanical Engineering,\\  The Johns Hopkins University, Baltimore, MD 21218, USA}

	\author{Cristian C. Lalescu}
	\affiliation{Max Planck Institute for Dynamics and Self-Organization, 37077 G\"{o}ttingen, Germany}

	\author{Michael Wilczek}
	\email{michael.wilczek@ds.mpg.de}
	\affiliation{Max Planck Institute for Dynamics and Self-Organization, 37077 G\"{o}ttingen, Germany}
				
\date{\today}

\begin{abstract}
We present an analysis of the Navier-Stokes equations based on a spatial filtering technique to elucidate the multi-scale nature of fully developed turbulence. In particular, the advection of a band-pass-filtered small-scale contribution by larger scales is considered, and rigorous upper bounds are established for the various dynamically active scales. The analytical predictions are confirmed with direct numerical simulation data. The results are discussed with respect to the establishment of effective large-scale equations valid for turbulent flows.
\end{abstract}


\maketitle

\section{Introduction}

The nonlinear and nonlocal nature of the incompressible Navier-Stokes equations gives rise to fluid turbulence as a  multi-scale phenomenon with a broad range of active spatial and temporal scales.
Nonlinear advection, for example, induces both a spatial translation as well as a distortion of the velocity field. Nonlocal pressure contributions, on the other hand, give rise to long-range interaction and therefore couple spatial scales in a non-trivial way. In view of the complex spatio-temporal dynamics, it is natural to ask whether dominant dynamical effects can be identified, at least when only a certain range of spatial scales is considered.

Traditional multi-scale analysis techniques usually employ Fourier decomposition to study the dynamics of Fourier coefficients (see, e.g., \cite{monin07book} for an extensive overview). In this formulation, the Navier-Stokes nonlinearities induce a coupling of Fourier modes which complicates the interpretation and rigorous analysis of the resulting strongly coupled set of equations. Spatial coarse-graining is an attractive alternative for decomposing the equations of motion into contributions from various scales \cite{leonard1974energy,germano1992turbulence}. With this technique, one can discern various scales of the dynamics in a real-space formulation. The coarse-graining approach has become very popular as a basis for large-eddy simulations (LES) \cite{Meneveau2000,Sagaut2006} but has also had a profound influence on the theoretical analysis of the multi-scale nature of turbulence. For example, using this framework, Eyink \cite{eyink2005locality,eyinknotes} showed that the energy transfer in the inertial range is scale-local, a concept which was extended using a scale-decomposition of the velocity into band-passed contributions \cite{eyinkAluie2009,aluie2009localness} and generalizes to other non-linear systems including magnetohydrodynamics \cite{aluie2010scale}.

If dominant dynamical contributions in turbulence can be identified, this opens the possibility to establish potentially much simpler effective equations of motion. The Tennekes-Kraichnan random sweeping hypothesis is an example of such simplified effective equations of motion. Introduced by Kraichnan as an ad hoc toy model to test the Direct Interaction Hypothesis \cite{kraichnan64pof}, the random sweeping hypothesis assumes that the small scales of the flow are approximately advected by the large scales passively, i.e. without significant dynamical feedback or distortion. Tennekes \cite{tennekes75jfm} used this argument to explain the failure of a na\"{i}ve application of Kolmogorov's phenomenology to Eulerian frequency spectra. While the random sweeping hypothesis unquestionably has its shortcomings and should only be regarded as a coarse approximation to the Navier-Stokes dynamics, it has proven to be useful in assessing corrections to Taylor's frozen eddy hypothesis in turbulence with mean flow \cite{lumley65pof,wyngaard77jas,george89afm} and in exploring the spectrum of turbulence in the presence of waves \cite{Lumley1983}. As discussed in \cite{chen89pfa}, the randomness of the large scales leads to temporal decorrelation. The random sweeping hypothesis has therefore found application in modeling space-time correlations of turbulent flows, for example within the framework of the so-called elliptic model \cite{he06pre,zhao09pre}, or more recently in the derivation of model wavenumber-frequency spectra \cite{wilczek12pre,wilczek15jfm}. For a review of the study of space-time correlations in turbulence, we refer the reader to \cite{wallace14tam}.

Further, the analysis of \cite{eyink2005locality,aluie2009localness,eyinknotes} has shown that the sub-grid stress modeled in LES is constituted primarily of contributions from a band of scales near the filter width. An understanding of the dynamical behavior in this band of scales is key for advancing turbulence modeling within the LES framework. The theoretical challenge of understanding the multi-scale nature of turbulence, developing models to capture the essence of its highly complex behavior, as well as pinpointing the reasons for surprisingly good performance seen for ad hoc models such as the random sweeping hypothesis constitute the main motivation for the current investigation.

The purpose of this paper is to explore in detail the extent to which small scale structures in the inertial range of high Reynolds number turbulence can be thought of as being dominated by large-scale sweeping, i.e. $(\partial_t + \bf{U} \cdot \nabla) \bv = 0$ where $\bf{U}$ and $\bv$ represent large and small scale velocities, respectively.  In this paper, we adopt an interpretation $\bv$ as a {band-passed} velocity field defined by successive spatial coarse-graining at different scales. To this end, we consider the dynamical evolution of a band of scales in the inertial range and establish the scaling behavior of the error from large-scale sweeping based on rigorous upper-bounds. Retaining only the dominant contribution in the limit of large scale-separation and large Reynolds number, we test the range of validity of such an approximation with the help of direct numerical simulations. Our analysis, as confirmed by numerical observations, establishes the rate at which dynamical terms in the (appropriately non-dimensionalized) equations of motion become negligible with decreasing filter width compared to kinematic sweeping by large scale motions. Further, we show that the time interval over which such an approximation holds can be predicted considering upper bounds for the time-integrated equation. 

Recently, an analysis of turbulent channel flow was performed in \cite{geng2015taylor} assessing the validity of Taylor's sweeping hypothesis in that context.   This work, as we discuss later on, is based on a term-by-term analysis of the exact equations of motion and is similar in spirit to the present work.  The primary difference is that their work is based on {a Reynolds decomposition of the Navier-Stokes equations} and relies on the presence of a large-scale mean flow whereas our work holds for flows without walls and uses spatial coarse-graining to distinguish large and small-scale motion.

The paper is structured as follows. The dynamical equations for the band-passed velocity field are derived in \S \ref{sec:filtering}. In \S \ref{sec:sweeping}, rigorous upper bound arguments are used to predict the scaling of dynamical terms compared with pure large-scale advection both instantaneously and in the time-integrated sense for high Reynolds number turbulence. This analysis is confirmed using results from direct numerical simulations of homogeneous isotropic turbulence {(see Appendix \ref{appendix:DNS} for details)}. Conclusions are drawn in \S \ref{sec:implications}.

\begin{figure}
    \includegraphics[width=\textwidth]{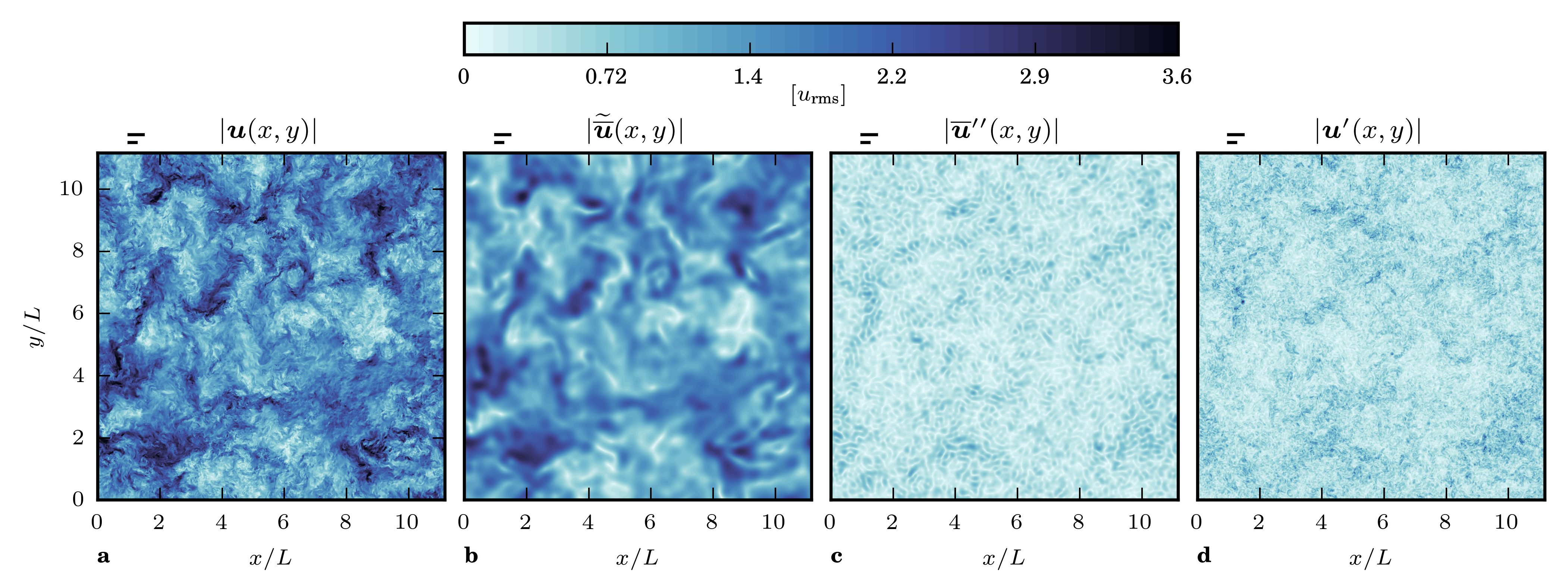}
    \caption{
    Visualizations of the amplitude of the different fields in \eqref{decomp}.
   \textbf{(a)}: full velocity, \textbf{(b)}: large-scale velocity, \textbf{(c)}: band-passed velocity,  {\textbf{(d)}}: remaining small-scale velocity.
    {The two small black segments in the top left of each panel correspond to the values of $\overline{\ell}$ and $\widetilde{\ell}$ used for these images, with $\tell/L\approx 0.64$ and $\overline{\ell} / \widetilde{\ell} = 0.50$.}
    \label{FigDecomp}}
\end{figure}

\section{Governing Equation of the band-passed velocity \label{sec:filtering}}

In this paper, we study properties of velocity fields $\bu$ which solve the Navier-Stokes equations on the torus $\mathbb{T}^d$:
\begin{align}\label{NSeqn}
\partial_t \bu + \bu \cdot \nabla \bu &=
    - \nabla p + \nu \Delta \bu + \bbf,\\
\nabla \cdot \bu &= 0,
\end{align}
where $\bbf$ is a large-scale body force.  We are interested in the dynamics of the velocity `between' two length scales $\oell,\tell$ with $\oell<\tell$.   To study this, following the approach of  Eyink \cite{eyink2005locality,eyinknotes},  Eyink \& Aluie \cite{eyinkAluie2009} 
we employ a coarse-graining or filtering procedure which allows us to decompose an arbitrary velocity field into three scale-localized pieces: (i) a large-scale component with scales $ > \tell$, (ii) a component with a band of scales between $\oell$ and $\tell$, and (iii) a small-scale component with scales $< \oell$. To make these precise, for each of the two scales $\oell,\tell$, we perform filtering at that scale with kernels $\oG= G_\oell$ and
$\widetilde{G} := G_\tell$ satisfying
\begin{equation}\label{filter}
G_\ell(\br):= \ell^d G(\br/\ell), \ \ \ \text{for } \ \ \ \ell= \oell,\tell,
\end{equation}
and hence
\begin{equation}
\nabla_\br G_\ell(\br)=\ell^{d-1} (\nabla G)(\br / \ell) :=  \ell^{-1} (\nabla G)_\ell (\br )
\label{eq:kernel_gradient_rescale}
\end{equation}
where the mother kernel $G$ is sufficiently smooth, rapidly decaying and normalized so that $\int d\br \ G(\br)=1$.  Coarse-graining the velocity corresponds to a low-pass filter, smoothing out fluctuations smaller than the given filter scale:
\begin{align}
\begin{split}
\overline{\bu}(\bx ,t) &= \int d\br \ \oG(\br) \ \bu(\bx  + \br,t).
\label{eq_filter1}
\end{split}
\end{align}
The residual field, $\bu' \equiv \bu - \overline{\bu}$, gives a concrete definition for the velocity with scales $< \oell$.  The gradient of the filtered field can be written in terms of the velocity increment $\delta\bu(\br;\bx,t) = \bu(\bx+\br,t) - \bu(\bx,t)$ using integration by parts to move the gradient operator onto the filter kernel,
\begin{equation}
\nabla \overline{\bu}(\bx)  = - \int d \br \ \nabla_\br \overline{G}(\br)  \bu(\bx  + \br)= -\frac{1}{\ell} \int d \br \ (\nabla G)_\ell (\br)  \delta\bu(\br;\bx),
\label{eq:filter_gradient}
\end{equation}
where we have used the property that $\int d \br \ \nabla_\br \overline{G}(\br) =0$ to subtract off $\int d \br \ \nabla_\br \overline{G}(\br)  \bu(\bx)$ in order to introduce the increment.  For a detailed pedagogical introduction to the machinery of coarse-graining, we refer the reader to \cite{eyinknotes}.  Filtering a second time with a larger filter width,
\begin{align}
\begin{split}\label{large-scale}
\widetilde{\overline{\bu}}(\bx ,t) = \int d\br \  \widetilde{\overline{G}} (\br)\  {\bu} (\bx  + \br,t),
\end{split}
\end{align}
where $\widetilde{\overline{G}} = \widetilde{G} * \overline{G}$, and $*$ denotes convolution. {Eq. \eqref{large-scale}}  defines the large-scale component in our decomposition with scales $>\tell$.
The velocity field band-passed between scales $\oell$ and $\tell$  is the residual of this second filtering operation, and can also be thought of as a filtered velocity increment:
\begin{align}
\begin{split}
\overline{\bu}''(\bx ,t) \equiv \overline{\bu} - \widetilde{\overline{\bu}} = \int d \br \  \overline{G}''(\br) \ \delta \bu(\br;\bx,t) ,
\label{eq:band-passed-velocity-increment}
\end{split}
\end{align}
where the filter kernel is $\overline{G}'' = (I-\widetilde{{G}} )*\overline{G} = \overline{G}-\widetilde{\overline{G}}$ . In this paper, the band-passed velocity field serves as a concrete mathematical manifestation of what is meant by a particular range of small-scale motions.

With these definitions in hand, the velocity decomposition can be stated precisely as:
\begin{equation}\label{decomp}
\bu =\widetilde{\overline{\bu}} + \overline{\bu}'' + \bu'.
\end{equation}
{Fig.~\ref{FigDecomp} shows a visualization of the decomposition; the large-scale velocity is essentially a smoothed version of the full velocity. The band-passed velocity shows a fine-scale structure with a characteristic length scale related to the length scales of the filtering. {The remaining small-scale velocity fluctuations are small in amplitude and exhibit less coherent structure relative to $\widetilde{\overline{\bold u }}$.}
{Clearly, the decomposition \eqref{decomp} of $\bu$ is purely passive and allows us to probe the interplay of scale-localized pieces of a given velocity field. As such, a major advantage of the coarse-graining approach is that all the results we obtain hold at the level of individual flow realizations without need to appeal to any subsequent statistical averaging.  It is important to note however that filtering introduces a degree of arbitrariness.  In particular, the precise definition of ``scale" as well as the structure of the velocity field in a scale-range will depend on the details of the chosen filter kernel.  Our analysis will apply to a broad class of possible filtering procedures and will focus on qualitative trends which are universally applicable independent of the choice of kernel $G$.} 

\begin{figure*}[t]
\includegraphics{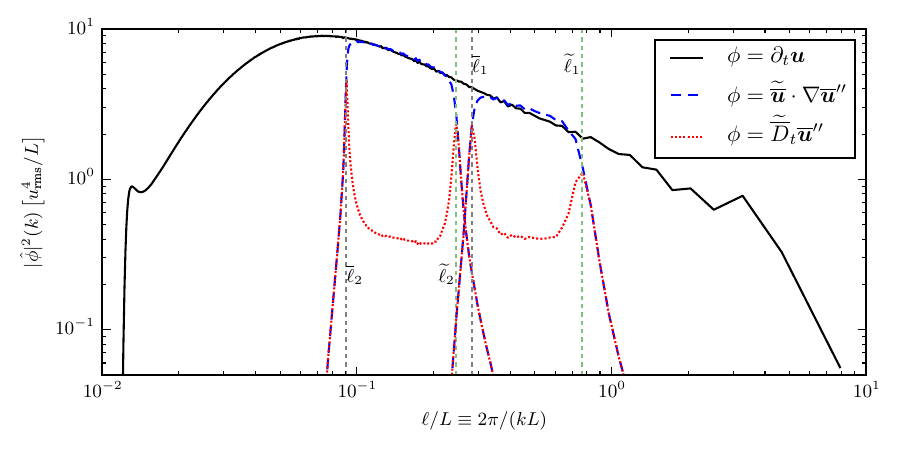}
\caption{
    Spectra (computed as an average over time slices) are shown for the Eulerian acceleration of full field $\partial_t \bs u$ (solid {black} line),  Lagrangian acceleration of band-passed velocity field $\widetilde{\overline{D}}_t\overline{\bu}'' $ (\textcolor{red}{red} dotted lines) and advection term $\widetilde{\overline{\bu}}\cdot \nabla\overline{\bu}'' $ (\textcolor{blue}{blue} dashed lines) for two specific bands $[ \smallfilter{\ell}_1,\bigfilter{\ell}_1] \approx [ 0.28L,0.76L] $ and $[ \smallfilter{\ell}_2,\bigfilter{\ell}_2]\approx [ 0.09L,0.25L] $.
    The dips of the dotted lines inside the intervals $[\smallfilter{\ell},
    \bigfilter{\ell}]$ indicate that the advective term nearly cancels the
    band-passed Eulerian acceleration.
    As the band is  pushed further into the inertial range, $[ \smallfilter{\ell}_1,\bigfilter{\ell}_1]\to [ \smallfilter{\ell}_2,\bigfilter{\ell}_2]$, the dip deepens, indicating that large-scale sweeping effects are becoming increasingly dominant.
}\label{fig2}
\end{figure*}

Filtering the incompressible Navier-Stokes equations sequentially at scales $\oell$ and $\tell$ and using the decomposition \eqref{decomp}, it is straightforward to obtain an evolution equation for the band-passed velocity field $\overline{\bu}''$ \eqref{eq:band-passed-velocity-increment}:
\begin{align}
\begin{split}
(\partial_t+\widetilde{\overline{\bu}}\cdot \nabla)\overline{\bu}''  &= \bbF \\ 
\text{with } \  &\bbF := - \overline{\bu}'' \cdot \nabla \widetilde{\overline{\bu}} - \overline{\bu}'' \cdot \nabla \overline{\bu}''  - \nabla \overline{p}'' + \nu\Delta\overline{\bu}'' \\
&\ \ \   \quad \ \ - \nabla \cdot \overline{\sigma}(\bu,\bu) - \nabla \cdot \widetilde{\overline{\sigma}}(\bu,\bu) + \overline{\bbf}''.
\end{split}
\label{eq:band-pass-Navier-Stokes}
\end{align}
Equation \eqref{eq:band-pass-Navier-Stokes} is stated succinctly as $\widetilde{\overline{D}}_t\overline{\bu}'' = \bbF$ where $\widetilde{\overline{D}}_t:=\partial_t+\widetilde{\overline{\bu}}\cdot \nabla$ is the material derivative along the large-scale flow.  The left-hand side $\widetilde{\overline{D}}_t\overline{\bu}'' $ of equation \eqref{eq:band-pass-Navier-Stokes} describes the {acceleration} of the band-passed scales by the large scale motions, and $\bbF$ represents the other forces. 
The first residual stress tensor, $\overline{\sigma}(\bu,\bu) = \overline{(\bu \otimes \bu)} - \overline{\bu} \otimes \overline{\bu}$, represents the net force on the $\overline{\bu}'' + \widetilde{\overline{\bu}} = \overline{\bu}$ component by the $\bu'$ component via nonlinear triadic interactions. This is the typical quantity to be modeled in LES with grid scale $\oell$ with $\bu'$ unresolved. In obtaining Eq. \eqref{eq:band-pass-Navier-Stokes} from successive filtering, the Germano identity \cite{germano1992turbulence} has been used, $\widetilde{\overline{\sigma}}(\bu,\bu) = \widetilde{\sigma}(\overline{\bu},\overline{\bu}) + \widetilde{\overline{\sigma}(\bu,\bu)}$, with the definition, $\widetilde{\overline{\sigma}}(\bu,\bu) = \widetilde{\overline{\bu \otimes \bu}} - \widetilde{\overline{\bu}} \otimes \widetilde{\overline{\bu}}$, where $\widetilde{\sigma}(\overline{\bu},\overline{\bu}) = \widetilde{\overline{\bu} \otimes \overline{\bu}} - \widetilde{\overline{\bu}} \otimes \widetilde{\overline{\bu}}$ is the Leonard stress \cite{leonard1974energy}. The resultant $\widetilde{\overline{\sigma}}(\bu,\bu)$ represents the net force on $\widetilde{\overline{\bu}}$ by $\overline{\bu}''+\bu'$ and roughly corresponds to the modeled quantity for an LES with grid scale $\tell$. The difference of these two stresses, then, represents the combined forcing on the $\overline{\bu}''$ components by both $\widetilde{\overline{\bu}}$ and $\bu'$ components via nonlinear triadic interactions. In addition to these residual stresses, the resolved force $-\overline{\bu}'' \cdot \nabla \widetilde{\overline{\bu}}$ represents distortion by the large-scale gradient. The other nonlinear resolved term, $-\overline{\bu}'' \cdot \nabla \overline{\bu}''$ represents the effect of $\overline{\bu}''$ advecting and distorting itself. The remaining terms are simply the band-passed right-hand side of the original Navier-Stokes equations.

The analysis of \eqref{eq:band-pass-Navier-Stokes}, both instantaneously and in a time-integrated sense, is the chief objective of this paper. 
 In Section \ref{sec:sweeping}, we investigate the strength of $\bbF$ in \eqref{eq:band-pass-Navier-Stokes} relative to pure advection of $\overline{\bu}''$ in high $Re$ turbulence as the scale $\tell$ is pushed deeper into the inertial range.  In this way, we evaluate the extent to which small-scales structures are advected by larger-scale coherent motions.

To initially explore why one might expect that advection of small-scale velocities by larger scales is a dominant effect, we investigate the spectrum of Eulerian accelerations $\partial_t \bs u$ from DNS data {of homogenous, isotropic turbulence} (described in more detail below) in Fig. \ref{fig2}. The spectrum, as expected, is very broad and shallow demonstrating the multitude of active scales present.  Using a spectral cut-off filter kernel, the band-passed Eulerian acceleration $\partial_{t}\overline{\bu}''$ (not shown) collapses by design with the Eulerian acceleration in the band and identically vanishes outside the band.  The important observation now is that the advection of the band passed velocity $\overline{\bu}''$ by the large scale velocity $\widetilde{\overline{\bu}}$ is also well localized in the band with very small contributions leaking outside the band. This shows that the Eulerian acceleration of the band-passed velocity is dominantly given by large-scale advection. To further confirm this, the sum $(\partial_{t}+\widetilde{\overline{\bu}}\cdot \nabla)\overline{\bu}''$ is shown in dotted lines, demonstrating a large degree of cancellation. The figure furthermore shows that the cancellations increase with scale separation from the large-scale advection velocity. These DNS observations, which extend those of Ref.\ \cite{Tsinober2001} to band-passed acceleration in the inertial range, {motivate our theoretical analysis of the advection of small-scale velocities by large-scale ones.}

\section{Analysis of Large-scale Sweeping Effects\label{sec:sweeping}}

\subsection{Scaling in high Reynolds number turbulence \label{sec:scaling}}
In high Reynolds number turbulent flows, the separation between the integral length scale $L$ and the viscous length scale $\eta = \nu^{3/4}\epsilon^{-1/4}$ becomes large, allowing for a significant inertial range of scales $\eta\ll \ell \ll L$.  In this range, it is well known that $p$th order (absolute) structure functions exhibit power-law behavior, 
\begin{equation}
S_p(\br) := \| \delta \bu(\br;\cdot) \|_{p}^{p} \sim   u_{ \rm rms}^p \left(\frac{|\br|}{L}\right)^{\zeta_p}, \quad \quad \eta\ll |\br| \ll L,
\label{eq:velocity-increment-scaling}
\end{equation}
where the increment is $\delta \bu(\br;\bx):=\bu(\bx+\br)- \bu(\bx) $ and  $\zeta_p$ is the scaling exponent (independent of $Re$). {Eq. \eqref{eq:velocity-increment-scaling} defines the structure function not by ensemble averaging but rather in terms of  spatial $L^p$ norms, $ \|f(\cdot) \|_p := \langle |f(\bx)|^p \rangle_x^{1/p} $ with $\langle \cdot \rangle_x$ denoting averaging in $\bx$}. 
When the relation \eqref{eq:velocity-increment-scaling} holds as an upper bound for all scales $|\br|$, it corresponds mathematically to Besov regularity (a global $L^p$ analogue of  H\"{o}lder continuity) with exponent $\sigma_p:=\zeta_p/p$ of the velocity field at high Reynolds numbers (see e.g. \cite{eyinknotes}).   We will work under the assumption that the solution to the Navier-Stokes equations satisfies this property with $0<\sigma_p\leq 1$, $1\leq p<\infty$ uniformly in Reynolds number.  
In Eq.~\eqref{eq:velocity-increment-scaling} and below we use the symbol $\sim$ whenever the relation should be interpreted as a scaling (i.e. upper and lower bound), rather than just an upper bound. Figure \ref{fig:scaling} (a) shows the structure function \eqref{eq:velocity-increment-scaling} for $p = 2,4,6$ (as examples) from DNS at $Re_\lambda \approx 437$.
A power-law scaling range with, e.g.,  $\zeta_4 = 1.27 \pm 0.01$ is observed for intermediate separation distances $30 \eta < |\br| < L$ indicated by vertical dashed red lines.   We use this specific definition of the inertial range. The exponents are estimated by computing the logarithmic derivatives of $S_p$ (see Fig. \ref{fig:scaling_exponents} and Appendix \ref{appendix:DNS} for details) and are within error of reported literature values (see e.g. \cite{anselmet84jfm}).

\begin{figure}
\includegraphics{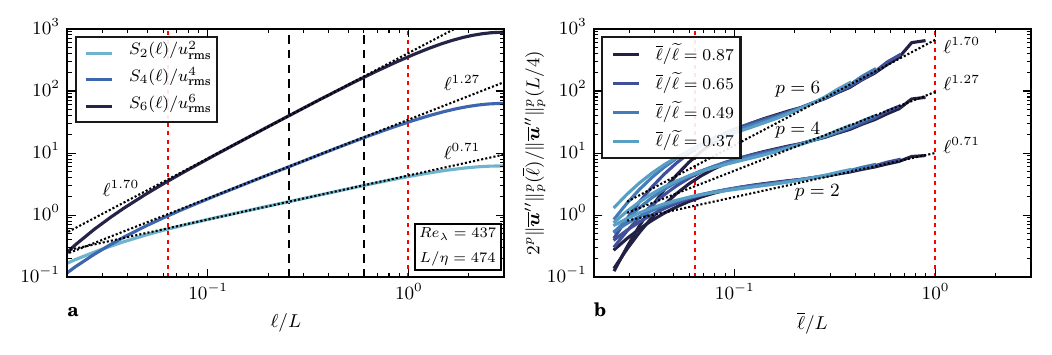}
\caption{\textbf{(a)}: Solid lines represent structure functions obtained from our DNS.
\textcolor{red}{Red} dashed vertical lines were placed at $30\eta$ and $L$, which define the extent of the `inertial range' in this paper. The {black} vertical dashed lines represent a typical
choice for the band of scales in the inertial range with $\oell\approx 0.25 L$ and $\tell=0.6 L$. Approximately a full decade of scaling is observed with exponents
$\zeta_2 = 0.71 \pm 0.01$,
$\zeta_4 = 1.27 \pm 0.01$,
$\zeta_6 = 1.70 \pm 0.02$.
\textbf{(b)}: Solid lines represent the $L^p$--norms for the band-passed velocity field \eqref{eq:band-passed-velocity-increment}, normalized by the value at $\overline{\ell} = L/4$ to emphasize the scaling region and multiplied by $2^p$ for clarity.
Approximate scaling is observed with the same exponents $\zeta_p$ calculated from the structure functions.}
\label{fig:scaling}
\end{figure}

Band-passed velocities can be related to increments using \eqref{eq:band-passed-velocity-increment}, and therefore \eqref{eq:velocity-increment-scaling} implies the following upper bound:
\begin{align}
\|\overline{\bu}''\|_{p} &= \left|\left| \int d \br \  \overline{G}_{\oell,\tell}''(\br) \ \delta \bu(\br;\cdot) \right|\right|_{p} \leq \int d \br \  |\overline{G}_{\oell,\tell}''(\br)| ~~ \|\delta \bu(\br;\cdot)\|_p\nonumber\\
&\leq u_{ \rm LS} \left(\frac{\tell}{L}\right)^{\sigma_p} \int d \rho \  |\overline{G}_{\oell,\tell}''(\rho)| ~  \rho^{\sigma_p} = \mathcal{O}\left( \left( \frac{\tell}{L}\right)^{\sigma_p}\right),  \quad \quad \eta \ll \oell < \tell\ll L,
\label{eq:band-passed-velocity-bound}
\end{align}
where $\rho = \br / \tell$.
 The `big-$\mathcal{O}$' notation indicates, as usual, inequality up to a constant independent of $\oell$ and $\tell$ separately, but which may depend on their ratio $\oell/\tell$ and the details of the mother filter $G$. Just as for structure functions, we expect that at large Reynolds numbers, \eqref{eq:band-passed-velocity-bound} would be reflected by scaling in an appropriate range.  Figure \ref{fig:scaling}(b) shows that, while not as prevalent, a {(limited)} power-law scaling range can be observed in our DNS data for $p = 2,4,6$ \footnote{Toward the smaller scales in the inertial range, a `bump'-like deviation is seen in the band-passed velocity scaling. This bump could be related to the well-known `spectral bump' deviation from $E(k) \sim k^{-5/3}$ observed in velocity spectra, since spectral cutoff filtering is used to generate Fig. \ref{fig:scaling}.}. {For higher Reynolds numbers, the scaling ranges for the structure functions grow in extent.  We expect that those for the  band-passed velocities also become longer and more clear.}

\subsection{Instantaneous Error in Large-scale Sweeping\label{sec:scaling_instant}}

{In this subsection,} we { directly evaluate the error in the instantaneous evolution equation \eqref{eq:band-pass-Navier-Stokes} when the right hand side terms are neglected. Upper bounds for the global error are considered using $L^p$ norms as in Eq. \eqref{eq:band-passed-velocity-bound} of the previous subsection. This provides insight as} to what extent the sweeping by large scales $\gtrsim \tell$ dominates the dynamics of the velocity within the band $[\oell,\tell]$, as that band is taken further into the inertial range, {i.e. taking the ratios $\oell/L, \tell/L$ to be increasingly small while maintaining that $\eta/\oell\ll 1$}.  Thus we will consider scales $\oell,\tell$ with fixed $\oell / \tell$.   We begin by deriving a bound on the $L^p$-norm of the Lagrangian derivative $\|\widetilde{\overline{D}}_t \overline{\bu}''   \|_p$ by the terms which balance it in the equations of motion \eqref{eq:band-pass-Navier-Stokes}.  {The Minkowski inequality allows us to bound the summation of all right hand side terms $\bbF$ in terms of the sum of their bounds,}
\begin{align}\nonumber
\|\widetilde{\overline{D}}_t \overline{\bu}''   \|_p = \|\bbF\|_p\  &\leq \  \|\overline{\bu}'' \cdot \nabla \widetilde{\overline{\bu}} \|_p  + \|\overline{\bu}'' \cdot \nabla \overline{\bu}''  \|_p  +\| \nabla \overline{p}'' \|_p  +\| \overline{\bbf}'' \|_p\nonumber \\
&\quad \quad \quad +\| \nu\Delta\overline{\bu}''  \|_p   +\|\nabla \cdot \overline{\sigma}(\bu,\bu)  \|_p + \|\nabla \cdot \widetilde{\overline{\sigma}}(\bu,\bu) \|_p. \label{MinkowskiBreak}
\end{align}
Each term of the right-hand side is bounded individually, assuming large Reynolds number $Re\gg 1$ and scale separation $\tell/L\ll 1$.  Such term-by-term analysis is similar in spirit to recent work \cite{geng2015taylor} assessing the validity of Taylor's hypothesis in turbulent channel flow, and follows very closely arguments found in \cite{eyink2005locality,eyinkAluie2009,eyink2015turbulent,eyinknotes}.  In particular, Eyink \cite{eyinknotes} performs an analogous computation for the filtered equations governing the coarse-grained velocity $\overline{\bu}_\ell$. { We next show in detail how upper bounds are derived for one of the terms in \eqref{MinkowskiBreak}. For the sake of brevity, the detailed calculations bounding other terms are given in Appendix \ref{app:instantaneous} and the resulting bounds are simply stated here in the main text.}

{As an example, we bound the first term on the right-hand side of \eqref{MinkowskiBreak}. The H\"{o}lder inequality  allows us to bound the product in terms of bounds on the two multiplicands,}
\begin{equation}
\|\overline{\bu}'' \cdot \nabla \widetilde{\overline{\bu}} \|_p \leq  \|\overline{\bu}'' \|_{q}\| \nabla \widetilde{\overline{\bu}} \|_{r}, \quad \quad \frac{1}{p} = \frac{1}{q} + \frac{1}{r}.
\label{eq:Holder}
\end{equation}
Although this choice is somewhat arbitrary, we set $q=r=2p$ for the purposes of demonstration in this paper.  Figure \ref{fig:scaling}(b) shows that for different values of $p$, $ \|\overline{\bu}''\|_{p}$ displays similar inertial range scaling properties so this choice is unlikely to have much influence, at least for relatively low moments.  
The first multiplicand in \eqref{eq:Holder} is bounded by \eqref{eq:band-passed-velocity-bound} and the second can be bounded using integration by parts, \eqref{eq:filter_gradient}, to move the gradient onto the filter kernel,
\begin{align}
\| \nabla \widetilde{\overline{\bu}} \|_r  &= \left|\left| \int d \br \ (\nabla \widetilde{\overline{G}})(\br)  \delta\bu(\br;\cdot) \right|\right|_r \leq \int d \br  |(\nabla \widetilde{\overline{G}})(\br)|  ~ \|\delta\bu(\br;\cdot)\|_r \nonumber\\
&\leq \frac{u_{ \rm LS}}{\tell} \left(\frac{\tell}{L}\right)^{\sigma_r} \int d \rho  |(\nabla G)(\rho)|  ~ \rho^{\sigma_r}  = \mathcal{O}\left( \left( \frac{\tell}{L}\right)^{\sigma_r - 1}\right),\quad \quad \eta \ll \oell < \tell\ll L,
\label{eq:bound_gradient}
\end{align}
where \eqref{eq:kernel_gradient_rescale} has been used in the scaling of the kernel gradient and \eqref{eq:velocity-increment-scaling} has been used to scale the norm of the increment.  Combining \eqref{eq:band-passed-velocity-bound} and \eqref{eq:bound_gradient} into \eqref{eq:Holder}, we obtain { the following upper bound for the first term in \eqref{MinkowskiBreak},}
\begin{equation}
\|\overline{\bu}'' \cdot \nabla \widetilde{\overline{\bu}}\|_p = \mathcal{O}\left(\left(\frac{\tell}{L}\right)^{2\sigma_{2p}-1}\right).
\end{equation}
{While this upper bound is shown in detail as an example, it is shown in Appendix \ref{app:instantaneous} that} a number of the other terms which comprise $\bbF$ are demonstrably the same order,
\begin{equation}\label{otherTerms}
\|\overline{\bu}'' \cdot \nabla \overline{\bu}''  \|_p, \  \| \nabla \overline{p}'' \|_p,\  \|\nabla \cdot \overline{\sigma}(\bu,\bu)  \|_p, \ \|\nabla \cdot \widetilde{\overline{\sigma}}(\bu,\bu) \|_p = \mathcal{O}\left(\left(\frac{\tell}{L}\right)^{2\sigma_{2p}-1}\right),
\end{equation}
which hold, again for scales $\oell,\tell$ in the range $\eta \ll \oell < \tell\ll L$.  The remaining two terms, $\| \overline{\bbf}'' \|_p$ and $\| \nu\Delta\overline{\bu}''  \|_p$, do satisfy the same upper bound as \eqref{otherTerms} but are actually considerably smaller asymptotically since we consider here smooth large-scale forcing schemes and high Reynolds numbers (small viscosity).   See Appendix \ref{app:instantaneous} for further details.  Combining these results, we obtain a bound on the large-scale sweeping term for any $p\geq 1$:
\begin{align}
\begin{split}
\|\widetilde{\overline{D}}_t \overline{\bu}'' \|_p = \|\bbF  \|_p &=  \mathcal{O} \left( \left(\frac{\tell}{L}\right)^{2\sigma_{2p}-1}\right),\quad \quad \quad  \eta \ll \oell < \tell\ll L.
\end{split}
\label{eq:sizeF}
\end{align}
By the same process{, i.e. using the H\"{o}lder inequality \eqref{eq:Holder} to separately bound multiplicands and integration by parts \eqref{eq:filter_gradient} to bound gradients,} the large-scale advection alone {can be shown to have a different upper bound,}
\begin{align}
\begin{split}
\| \widetilde{\overline{\bu}}\cdot \nabla\overline{\bu}''\|_{p}&=  \mathcal{O} \left( \left(\frac{\tell}{L}\right)^{\sigma_{2p}-1}\right) ,\quad \quad\quad\quad\quad \eta \ll \oell < \tell\ll L.
\end{split}
\label{eq:band-pass-acceleration-advective}
\end{align}
{Finally, moving the large-scale advection to the right-hand side of \eqref{eq:band-pass-Navier-Stokes} before applying the Minkowski inequality, an upper bound for the Eulerian time derivative can be computed,
$
\| \partial_{t}\overline{\bu}''\|_{p} \leq \| \widetilde{\overline{\bu}}\cdot \nabla\overline{\bu}''\|_{p} + \|\bbF  \|_p.
$
Substituting \eqref{eq:sizeF} and \eqref{eq:band-pass-acceleration-advective} and recalling our assumption that $0<\sigma_{p} \leq 1$ for all $p\geq1$, we obtain the upper bound:}
\begin{align}
\begin{split}
\| \partial_{t}\overline{\bu}''\|_{p}&= \mathcal{O} \left( \left(\frac{\tell}{L}\right)^{\sigma_{2p}-1}\right) ,\quad \quad\quad\quad\quad\quad\quad \eta \ll \oell < \tell\ll L.
\end{split}
\label{eq:band-pass-acceleration-Eulerian}
\end{align}
 {It is worth emphasizing that \eqref{eq:sizeF}--\eqref{eq:band-pass-acceleration-Eulerian} constitute rigorous upper-bounds, which follow from an exact coarse-graining analysis of the Navier-Stokes equations.  
We see from \eqref{eq:sizeF} that the bound on the deviation from pure large-scale advection quantified by the magnitude of $\bbF$ grows as $\tell$ is taken farther into the inertial range ($\tell/L\to 0$) provided that the flow is sufficiently rough, i.e. $\sigma_{p}< 1/2$ for $p\geq 2$ \footnote{This bound on the scaling exponent $\sigma_p$ is consistent with reported values in the literature, as well as our own data since $\sigma_2= 0.355\pm 0.005$ and $\sigma_p$ is a decreasing function of $p$, see \cite{eyinknotes}.}. 
Moreover, the bounds on the pure large-scale advection and the Eulerian acceleration terms also both grow  as $\tell/L$ decrease even for nearly smooth flows with $\sigma_p<1$.  

 The individual terms of \eqref{eq:band-pass-Navier-Stokes} involve spatial/temporal derivatives of the fine-grained velocity field, regulated by spatial filtering.  The bounds \eqref{eq:sizeF}--\eqref{eq:band-pass-acceleration-Eulerian}  diverge for rough turbulent flow fields because these derivatives are dissipation range objects which do not exist in the naive sense as $Re\to\infty$ but rather only as distributions.  This suggests studying large-scale sweeping $\widetilde{\overline{D}}_t \overline{\bu}'' $ in a renormalized system of units where the large-scale advection term $\widetilde{\overline{\bu}}\cdot \nabla\overline{\bu}''$ is $\mathcal{O}(1)$  by design for all scales $\oell,\tell$ in the inertial range,  thus allowing us to probe the relative strength of sweeping compared to the other physical processes.

 To perform a meaningful non-dimensionalization, one must recognize that there are multiple velocity scales in the problem. In particular, there is the typical size $u_{ \rm LS}$ of the large-scale sweeping velocity $\widetilde{\overline{\bu}}$ and there is the fluctuation velocity $\delta u(\tell)$ which measures the typical size of the band-passed field $\overline{\bu}''$.  We define these scales precisely in terms of spatial $L^q$ norms for a given $1\leq q<\infty$: 
\begin{equation}\label{scales}
 u_{ \rm LS} := \|\widetilde{\overline{\bu}}\|_{q}, \ \ \ \ \delta u(\tell):=\|\overline{\bu}''\|_{q}
\end{equation}
where these are computed, for example, at a fixed time or as a average over different times.  The velocity scales \eqref{scales} can be used to precisely define ``sweeping" $t_\tell$  and  local eddy turnover $\tau_\tell$ time scales: 
\begin{equation}\label{timescales}
t_\tell :=\tell/u_{ \rm LS}, \ \ \ \ \tau_\tell:=\tell/\delta u(\tell).
\end{equation}
Assuming that the band-passed field displays scaling (as discussed below Eq.~\eqref{eq:band-passed-velocity-bound} and is corroborated by DNS data in Figure \ref{fig:scaling}), and that there is weak scale-dependence for the $L^q$-norms the large-scale field \eqref{large-scale}, then the scales used for non-dimensionalization \eqref{scales} and \eqref{timescales} behave according to:
\begin{equation}\label{scalesScaling}
 u_{ \rm LS} \sim \mathcal{O}\left(1\right),\ \ \ \ \delta u(\tell)\sim \mathcal{O}\left( \left( \frac{\tell}{L}\right)^{\sigma_q}\right),\ \ \ \  t_\tell \sim \mathcal{O}\left( \frac{\tell}{L}\right), \ \ \ \  \tau_\tell \sim \mathcal{O}\left( \left( \frac{\tell}{L}\right)^{1-\sigma_q}\right).
\end{equation}
 We form a non-dimensional equation using  the following rescaling $\widetilde{\overline{\bu}}^*:= \widetilde{\overline{\bu}}/ u_{ \rm LS}$ and $\overline{\bu}''^*:= \overline{\bu}''/\delta u(\tell)$, the length scale $\bx^*:=\bx/\tell$ and the ``sweeping" timescale $t^*:= t/t_\tell$. Equation \eqref{eq:band-pass-Navier-Stokes} in these variables becomes
\begin{align}
\begin{split}
(\partial_{t^*}+\widetilde{\overline{\bu}}^*\cdot \nabla^*)\overline{\bu}''^*  &= {\bold F}^* 
\end{split}
\label{eq:band-pass-Navier-Stokes-nondim}
\end{align}
where ${\bold F}^*:={\bbF}/(u_{ \rm LS}/\tau_\tell)$. Together with the scalings \eqref{scalesScaling},  non-dimensionalizing the bounds \eqref{eq:sizeF}--\eqref{eq:band-pass-acceleration-Eulerian}  yield:
\begin{align}
\begin{split}
\| \partial_{t^*}\overline{\bu}''^*\|_{p} =  \mathcal{O}(1), \quad \quad \| \widetilde{\overline{\bu}^*}\cdot \nabla^*\overline{\bu}''^*\|_{p} =  \mathcal{O}(1), \quad \quad\|\widetilde{\overline{D}}_t^* \overline{\bu}''^*   \|_p = \mathcal{O} \left( \left(\frac{\tell}{L}\right)^{\sigma_{2p}}\right),\quad \quad \quad  \eta \ll \oell < \tell\ll L.
\end{split}
\label{eq:band-pass-acceleration-Lagrangian}
\end{align}
 In the definition \eqref{scales}, $q$ is arbitrary; however in deriving \eqref{eq:band-pass-acceleration-Lagrangian}, we have set $q=2p$ since this is the unique choice for which the derived bound on the large-scale advective term is $\mathcal{O}(1)$. Now, the non-dimensionalized  term $\widetilde{\overline{D}}_t^* \overline{\bu}''^*$ \eqref{eq:band-pass-acceleration-Lagrangian} vanishes (in a spatial $L^p$ sense) as $\tell/L\to 0$ provided only that $\sigma_{p}> 0$ for $p\geq 2$, indicating that sweeping becomes an increasingly dominant effect as the band $[\oell,\tell]$ is taken further into the inertial range.  Of course, the smoother the velocity (the closer the exponent $\sigma_{2p}$ is to unity), the more dominant the large-scale sweeping effect.  This is expected as small-scale irregularities will tend to distort coherent motion of the fluid.  However, the result \eqref{eq:band-pass-acceleration-Lagrangian} shows that even for turbulent velocity fields which are spatially irregular (satisfying only that $\sigma_{2p}> 0$), large-scale sweeping is the single dominant dynamical effect on small-scale motions.   Essentially, our analysis gives a precise interpretation for the statement that, for an appropriate range of scales and properly non-dimensionalized, `small-scale fluctuations are principally swept by the large-scale flow'.  It should be emphasized that these results follow from an exact mathematical analysis of the incompressible Navier-Stokes equations and not dimensional arguments. We further confirm these observations by studying time-integrated errors in the section \ref{sec:scaling_cumul}.   

\begin{figure}
\includegraphics{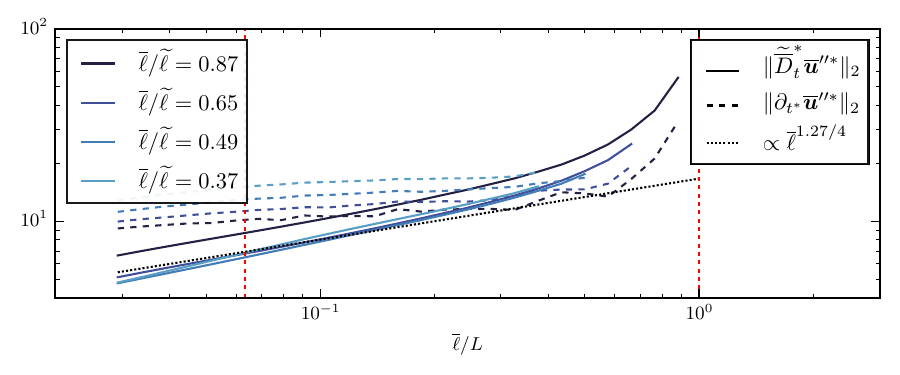}
\caption{Here we quantify the qualitative results presented in Fig. \ref{fig2}.  \textcolor{red}{Red} dashed vertical lines indicate the inertial range $[30\eta,L]$ empirically defined by Fig. \ref{fig:scaling}. As bands of scales $[\oell,\tell]$ for different ratios $\oell/\tell$ are pushed further into the inertial range, the instantaneous approximation
$(\partial_{t^*}+\widetilde{\overline{\bu}}^*\cdot \nabla^*)\overline{\bu}''^*=0$ is seen to work increasingly well.  The decrease with $\tell/L$ is observed to be slightly steeper than, but consistent with, that of the upper bound \eqref{eq:band-pass-acceleration-Lagrangian}.
On the other hand, the Eulerian time derivative $\partial_{t^*}\overline{\bu}''^*$ alone remains nearly flat and is also consistent with \eqref{eq:band-pass-acceleration-Lagrangian}. }
\label{fig:band-pass-acceleration-scale}
\end{figure}

Although we have no compelling theoretical argument to show that the upper bounds \eqref{eq:band-pass-acceleration-Lagrangian} hold as power-law scaling laws within the inertial range, it is interesting to quantify the extent to which they do.  For the purpose of illustration, we compute $2$-norms in \eqref{eq:band-pass-acceleration-Lagrangian}, i.e., $p = 2$ with our DNS data.  Figure \ref{fig:band-pass-acceleration-scale} shows separately the dependence of  Eulerian and Lagrangian (following the coarse-grained velocity field) derivative on filter scale $\oell/L$ for different $\oell / \tell$. The numerical data shows that the Lagrangian acceleration of band-passed velocity drops off more steeply than the rate given  \eqref{eq:band-pass-acceleration-Lagrangian}, particularly for $\oell \approx L$ but with better agreement for $\oell \ll L$.  This suggests that the upper bound is asymptotically larger than the actual magnitude of $\|\widetilde{\overline{D}}_t^* \overline{\bu}''^*\|_2$ as $\tell/L\to 0$, indicating that the sweeping is an even stronger effect than the analysis indicates.   In contrast, the Eulerian acceleration of the band-passed velocity is relatively constant but with a slight decrease in this range, again consistent with the upper bound \eqref{eq:band-pass-acceleration-Lagrangian}. The increasing gap between these two accelerations as $\oell/L$ decreases shown in Figure \ref{fig:band-pass-acceleration-scale} signifies the increasing dominance (in the sense of $L^p$ norms) of the large-scale sweeping.

{It should be reiterated that the coarse-graining analysis which we apply here is applicable to nearly any flow in domains without boundary, not just the homogenous isotropic turbulence chosen here as a case study.  The basic assumption on the flow that we make is that the velocity has finite moments ($L^p$--norms) and that structure functions satisfy  \eqref{eq:velocity-increment-scaling}  as upper bounds  for some positive scaling exponents $\zeta_p>0$ independent of $Re$. However, the extent to which the bounds \eqref{eq:velocity-increment-scaling} for structure functions and \eqref{eq:band-passed-velocity-bound} for band-passed fields hold as scalings in an appropriate range may vary from flow to flow.  Therefore, \eqref{eq:band-pass-acceleration-Lagrangian} may also hold correspondingly better or worse as a scaling.}

\subsection{Cumulative Error in Passive Advection Approximation\label{sec:scaling_cumul}}

To further assess the relative dominance of large-scale advection, as well as the associated time scales, we examine the error committed by replacing \eqref{eq:band-pass-Navier-Stokes} with,
\begin{align}\label{eq:band-pass-passive-advection}
\partial_{t} \bv + \widetilde{\overline{\bu}}\cdot \nabla \bv &=0, 
\end{align}
where $\bv$ is passively advected.   
We wish to quantify the extent to which \eqref{eq:band-pass-passive-advection}  is a faithful approximation for the band-passed velocity as an evolution equation when integrated for a finite time.  Clearly \eqref{eq:band-pass-passive-advection} is an idealized model which neglects many important physical processes and, as such, we do not expect $\overline{\bu}''$ and $\bv$ to agree for very long.   Nevertheless, we argue that the solution $\bv$ approximates the band-passed velocity $\overline{\bu}''$ in a spatially global sense for times set by the scales $t_\tell$ and $\tau_\tell$ defined by \eqref{timescales}.  

 To investigate this, we consider $\overline{\bu}''^*$ the solution of \eqref{eq:band-pass-Navier-Stokes-nondim} for initial data $\overline{\bu}''^*(\bx,0) = \overline{\bu}''^*_0(\bx)$ and let $\bv =\bv(\bx,t)$ satisfy the passive vector advection equation \eqref{eq:band-pass-passive-advection} for the same initial data.  The error $\bw:= \overline{\bu}''-\bv$ then satisfies:
\begin{align}\label{deviationEqn}
(\partial_{t}+\widetilde{\overline{\bu}}\cdot \nabla)\bw  = \bbF ,  \ \ \ \ \ \bw(\bx,0) = 0,
\end{align}
where the $\bbF$ is given by \eqref{eq:band-pass-Navier-Stokes}. Introducing Lagrangian tracers in the large-scale filtered field:
\begin{align}\label{eq:filteredlagrangianparticles}
\frac{d}{dt} \otX_t(\bx) =  \otu(\otX_t(\bx), t),  \ \ \ \otX_0(\bx)= \bx, 
\end{align}
we can integrate equation \eqref{deviationEqn} to find: 
\begin{align}
\overline{\bu}''(\bx,t)-\bv(\bx,t)  =  \int_0^tds \ \bbF (\otX_s^{-1}(\bx),s).
\end{align}
Since $\otu$ is a smooth incompressible field, the associated Lagrangian flow $\otX_t$ is volume preserving.  Therefore,
\begin{align}\label{timebound}
\|\overline{\bu}''(t)-\bv (t)\|_p \leq  \int_0^tds \ \|\bbF (\cdot,s)\|_p\leq \sup_{0 \leq s\leq t}\|\bbF (\cdot,s)\|_p t.
\end{align}
For quasi-stationary homogeneous turbulence,  $\|\bbF (\cdot,t)\|_p$ is nearly constant-in-time and so the first upper bound is likely to be a near-equality.  Non-dimensionalizing velocities by $\delta u(\tell)$ \eqref{scales} and times by $t_\tell$  \eqref{timescales}, we find
\begin{align}\label{cumulver1}
\|\overline{\bu}''^*(t)-\bv^* (t)\|_p \leq  \sup_{0 \leq s\leq t}\|\bbF^* (\cdot,s)\|_p \left(\frac{t}{t_\tell}\right) = \mathcal{O} \left( \left(\frac{\tell}{L}\right)^{\sigma_{2p}}\right) \times \left(\frac{t}{t_\tell}\right), \quad \quad\quad \eta \ll \oell < \tell\ll L.
\end{align}
  Thus, for times on the order of the sweeping timescale $t_\tell$, we see that the error relative to a typical fluctuation $\delta u(\tell)$ decreases in accord with the instantaneous bound of Eq.~\eqref{eq:band-pass-acceleration-Lagrangian}.    Alternatively, the inequality \eqref{cumulver1} can be restated as
\begin{align}
\|\overline{\bu}''^*(t)-\bv^* (t)\|_p = \mathcal{O}(1)\times  \left(\frac{t}{\tau_\tell}\right),\quad \quad\quad\quad\quad \eta \ll \oell < \tell\ll L.
\label{eq:wscale}
\end{align}
 This shows that the error relative to $\delta u(\tell)$ deviates in a controlled way from the true evolution for times of order the eddy-turnover time at the outer length-scale of the fluctuation, $\tau_\tell$.  
 
 Thus, our analysis of the `passive advection approximation' confirms that $t_\tell$ and $\tau_\tell$ are the relevant time scales for large-scale sweeping.   For times on the order $t_\tell$, large-scale sweeping becomes increasingly dominant as $\tell\ll L$ is taken further into the inertial range, albeit for shorter times and smaller fluctuations  $t_\tell\sim\tell/L$ and $\delta u(\tell)\sim(\tell/L)^{\sigma_{p}}$ (see Eq. \eqref{scalesScaling}). On the other hand, for asymptotically longer times of the order of the local-eddy turnover time $\tau_\tell\sim(\tell/L)^{1-\sigma_{2p}}$, the passive approximation does not become more accurate but instead the errors remain order unity and controlled.  {In this precise sense, one can view $\tau_\tell$ as the time-scale relevant for the large-scale Lagrangian dynamics, since errors away from large-scale transport cannot accumulate in this time. These conclusions have previously been drawn on the level of intuitive physical arguments \cite{tennekes1972first}, but our analysis puts these results on a firmer mathematical footing, lending more support to the classical Tennekes-Kraichnan random sweeping picture  \cite{kraichnan64pof,tennekes75jfm}.}

\begin{figure}
\includegraphics{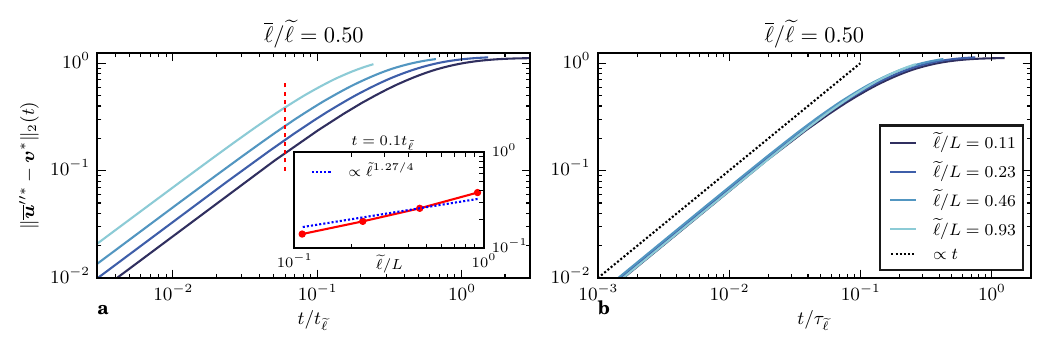}
\caption{
(a) and (b): Solid lines represent $\|\overline{\bu}''^*(t)-\bv^* (t)\|_2$ plotted for times of order $t_\tell$ and $\tau_\tell$ respectively at one fixed value of $\oell / \tell = 0.5$ and four values of $\tell/L= 0.11, 0.23, 0.46, 0.93$. Linear growth of errors relative to the passive advection approximation are observed. The inset of panel (a) illustrates that, for fixed $t/t_\tell$, the error (\textcolor{red}{red} solid line) decreases with $\tell/L$ nearly as a power law with a steeper slope than that predicted by the bound \eqref{cumulver1} (\textcolor{blue}{blue} dashed line).  On the other hand, panel (b) shows that the error is independent of $\tell/L$ at fixed $t/\tau_\tell$, consistent with the bound \eqref{eq:wscale}. }
\label{fig:band-pass-acceleration-scale2}
\end{figure}

We tested the predictions \eqref{cumulver1} and \eqref{eq:wscale} numerically with our DNS data. To do this, we solve \eqref{eq:band-pass-passive-advection} with the method of characteristics, expressing $\bv(\bx,t)=\overline{\bu}_0''(\otX_t^{-1}(\bx))$.  We then evaluated, at a given time $t$, the error \eqref{timebound}:
\begin{align}\label{lagTermsEval}
\|\overline{\bu}''(\cdot,t)-\bv (\cdot,t)\|_p = \|\overline{\bu}''(\cdot,t)-\overline{\bu}_0''(\otX_t^{-1}(\cdot))\|_p =  \|\overline{\bu}''(\otX_t(\cdot),t)-\overline{\bu}_0''(\cdot)\|_p,
\end{align}
where we have used the fact that the flow $\otX_t$ preserves volume to change variables inside the $L^p$ norm.
To evaluate the error given by the final expression in \eqref{lagTermsEval}, we placed particles homogeneously in the flow and advected them solely using the large-scale velocity according to \eqref{eq:filteredlagrangianparticles} (see Appendix \ref{appendix:DNS} for details on the numerical integration scheme employed).
The band-passed velocity was then sampled along the tracer trajectories and, after non-dimensionalizing, the error  \eqref{cumulver1} was computed.

 Figure \ref{fig:band-pass-acceleration-scale2} (a) and (b) show the error $\|\overline{\bu}''^*(t)-\bv^* (t)\|_2$ over times which extend to $t_\tell$ and $\tau_\tell$ respectively.  We chose values of $\oell, \tell$ so that each band $[\oell,\tell]$ falls into the empirical inertial range $[30\eta, L]$ established in Section \ref{sec:scaling}. Both plots show an initial period of linear growth of the error with slope predicted by \eqref{timebound} for quasi-stationary turbulence.  This behavior is followed by a leveling off of the curves at larger times due to a saturation of error as the bandpass velocity become uncorrelated with its initial value. 
 Further, Figure \ref{fig:band-pass-acceleration-scale2} (a) shows that, for times on the order of $t_\tell$, there is a decrease with $\tell/L$ consistent with the upper bound given in  \eqref{cumulver1} as $\tell$ is taken further in the inertial range.   On the other hand, Figure \ref{fig:band-pass-acceleration-scale2} (b)  shows that for times on the order of $\tau_\tell$, the error displays a very weak dependence on the scales $\oell,\tell$, as predicted by \eqref{eq:wscale}, confirming our theoretical predictions.

\section{Conclusions\label{sec:implications}}

In this paper, we employed a spatial filtering approach to study the extent to which sweeping by a large-scale flow is a dominant effect on the Navier-Stokes dynamics of small-scale velocities.
In order to investigate well-defined scales in the inertial range, we considered the advection of the band-passed velocity by the large-scale velocity.
In Section \ref{sec:filtering}, an equation of a band-passed velocity was derived, and in Section \ref{sec:sweeping} the magnitude of the individual terms was rigorously estimated in the sense of $L^p$ norms under the assumption of a high Reynolds number flow exhibiting scaling of the Eulerian structure functions (discussed in Section \ref{sec:scaling}).  We then exposed the spatial/temporal scales relevant for large-scale sweeping.  In particular, we showed that if an appropriate and natural nondimensionalization is employed, the advection of the band-passed velocities by the large-scale velocity is increasingly the dominant effect as the band is pushed further in the inertial range, lending support to the classical Tennekes-Kraichnan random sweeping hypothesis. This was demonstrated both instantaneously (Section \ref{sec:scaling_instant}) and in a time-integrated sense (Section \ref{sec:scaling_cumul}), and our detailed scaling predictions were found to be in good agreement with DNS simulation data {of homogenous, isotropic turbulence}.

We have to stress that the results presented here are cast in terms of $L^p$ norms. The passive advection approximation is therefore expected to only hold in a spatial average sense. Locally, i.e.~point by point, the approximation may be considerably less faithful. The (limited) validity of the approximation globally may partially explain the generally good performance of random-sweeping based models for the space-time correlations in turbulent flows such as those discussed in \cite{wilczek12pre}. In theses models, the temporal decorrelation of turbulence is assumed to be governed by large-scale random sweeping effects. As space-time correlations are defined in terms of statistical averages, which for the evaluation of numerical or experimental data are usually replaced by spatial and temporal averages, the results presented here apply.

Other, more general insights into the multi-scale behavior of turbulence can be gleaned from these results as well.  
 For instance, the time scale $\tau_\tell$ at which sweeping approximation breaks down, i.e., the time scale for which band-passed velocities are correlated, highlights a limitation for applying the passive advection approximation, for example, in developing models for the wavenumber-frequency spectra. In particular, this analysis suggests that models using the passive advection approximation will not be as reliable in the range of low frequencies.  Moreover, the decorrelation time scale $\tau_\tell$, can be seen as a guide for the choice of the time scale for Lagrangian averaging in Lagrangian dynamic sub-grid models \cite{Meneveau1996,Bou-Zeid2005}. In that model, the relevant quantities (dominated within a band of scales near the filter width \cite{eyink2005locality,aluie2009localness,eyinknotes}) must be averaged long enough to provide sufficient smoothing for numerical stability without losing significant local information, i.e., over a time scale not much smaller nor much larger than $\tau_\oell$, which is borne out in the testing of the model.

In conclusion, our results exemplify that in certain limits, simplified effective equations of motion can be established for the properties of small-scale features in turbulent flows. Of course, this is a well-known statement and this basic physical idea has been used in previous modeling approaches, e.g., space-time correlation models \cite{wilczek12pre}. The present work, however, provides a rigorous mathematical basis for this understanding with precisely defined quantities, exemplifying the theoretical utility of the filtering approach for deciphering turbulent dynamics.
\vspace{-4mm}

\acknowledgements
\vspace{-4mm}

Computations were performed on the clusters of the Max Planck Computing and Data Facility.
C.C.L.~and M.W.~are supported by the Max Planck Society. M.W.~gratefully acknowledges support by the Priority Program SPP 1881 Turbulent Superstructures of the Deutsche Forschungsgemeinschaft. P.L.J. was supported by a Graduate Research Fellowship from the National Science Foundation (DGE-1232825).

\appendix

\section{Detailed Bounds\label{app:instantaneous}}

Here we complete the derivation of Eq. \eqref{eq:band-pass-acceleration-Lagrangian} by bounding individual terms in the inequality \eqref{MinkowskiBreak} of the main text.  The results and methods follow closely a similar term-by-term estimation of the large-scale equations for $\overline{\bu}_\ell$ given in Chapter IIb of \cite{eyinknotes}.  By the arguments given in the main text by Eqns. \eqref{eq:Holder}, \eqref{eq:bound_gradient}, we estimate:
\begin{align}
\begin{split}
 \|\overline{\bu}'' \cdot \nabla \widetilde{\overline{\bu}} \|_p                                            & \leq    \|\overline{\bu}'' \|_{2p}\| \nabla \widetilde{\overline{\bu}} \|_{2p}  = \mathcal{O}\left( \left( \frac{\tell}{L}\right)^{2\sigma_{2p} - 1}\right),\\
 \|\overline{\bu}'' \cdot \nabla \overline{\bu}''  \|_p      & \leq    \|\overline{\bu}'' \|_{2p}\| \nabla \overline{\bu}'' \|_{2p}      = \mathcal{O}\left( \left( \frac{\tell}{L}\right)^{2\sigma_{2p} - 1}\right).
\end{split}
\end{align}
{Note that since $\tell$ and $\oell$ are a fixed multiple of each-other, we need not distinguish between them in the upper bounds we obtain. In fact, for all terms in $\bbF$ except the viscous term (discussed below), we can obtain rigorous upper bounds which are valid for arbitrarily large $Re$ which hold directly for the fine grained field ${\bu}':=\bu- \overline{\bu}$ without the need for band-passing.}
Next, we bound the contribution from the turbulent stress using the identity:
\begin{align}
\nabla \cdot \overline{\sigma}_\ell (\bu,\bu) &= -\frac{1}{\ell}\int d \br \ (\partial_i G)_\ell(\br) \delta u_i(\br;\bx) \delta u_j(\br;\bx)  +  \frac{1}{\ell}\int d \br \ (\partial_iG)_\ell(\br) \delta u_j(\br;\bx)  \int d \br \ G_\ell(\br)\delta u_i(\br;\bx).
\end{align}
See Appendix B of \cite{eyink2015turbulent}  or Chapter IIb of \cite{eyinknotes} for a detailed derivation and discussion.  Using H\"{o}lder's inequality, a bound on the stress in terms of $L^p$ norms of the increments:
\begin{align}\nonumber
\|\nabla \cdot \overline{\sigma}_\ell (\bu,\bu) \|_p &\leq \frac{1}{\ell} \int d \br \ |\nabla G_\ell(\br)| \|\delta \bu(\br;\cdot)\|_{2p}\|\delta \bu(\br;\cdot)\|_{2p}\\
&  \quad +  \frac{1}{\ell}\iint d \br d \br' \ |\nabla G_\ell(\br)|  | G_\ell(\br')| \|\delta \bu(\br;\cdot )\|_{2p} \|\delta \bu(\br';\cdot)\|_{2p} .
\end{align}
Using the above, we deduce that:
\begin{align}
\begin{split}
  \|\nabla \cdot \overline{\sigma}(\bu,\bu)  \|_p, \ \|\nabla \cdot \widetilde{\overline{\sigma}}(\bu,\bu) \|_p  =\mathcal{O}\left( \left( \frac{\tell}{L}\right)^{2\sigma_{2p} - 1}\right).
\end{split}
\end{align}
Estimating the contribution of the pressure gradient is more subtle.  First, it is straightforward to obtain
\begin{align}
\| \nabla \overline{p}'' \|_p  &\leq \frac{1}{\oell} \int d \br \ |(\nabla \overline{G}'')(\br)|\| \delta {p}(\br;\cdot)\|_{p}.
\end{align}
{Estimates for pressure increments have been derived with the Kolmogorov 1941 (K41) theory, see e.g. \cite{obukhov1949pressure,batchelor1951pressure}.}  However, to rigorously bound $\| \delta {p}(\br;\cdot)\|_{p}$ {without appealing to K41 arguments}, it is necessary to appeal to additional mathematical theory.  In particular, it can be shown from the pressure-Poisson equation $-\Delta p = \partial_i u_j\partial_j u_i$ that if the velocity structure function satisfies  \eqref{eq:velocity-increment-scaling} for all scales $|\br|$ and $Re$, then the pressure structure function obeys:
\begin{align}
\| \delta {p}(\br;\cdot)\|_{p}&= \mathcal{O}\left(   \left(\frac{|\br|}{L}\right)^{2\sigma_{2p}}\right),
\end{align}
with $\sigma_{2p} :=  \zeta_{2p}/2p$ where $\zeta_{2p}$ is the scaling exponent for $2p$-structure function of the velocity field.   This can be deduced from elliptic regularity theory for Calder\'{o}n--Zygmund operators.  A clear discussion of this issue, as well as further references, can be found in Chapter IIb of \cite{eyinknotes} {and Section 4.2.4. of \cite{triebel2006theory}}.  Thus, we obtain the bound:
\begin{align}
\|\nabla  \overline{p}'' \|_p &= \mathcal{O}\left( \left( \frac{\tell}{L}\right)^{2\sigma_{2p} - 1}\right).
\end{align}
The contribution of the forcing is estimated using the identity \eqref{eq:band-passed-velocity-increment} for band-passed objects:
\begin{align}
\begin{split}
\|\overline{\bbf}'' \|_p                &  \leq \int d\br \ |\overline{G}''(\br)|\   \|\delta f(\br;\cdot)\|_p    = \|\nabla f\|_\infty\times \mathcal{O}\left(  \frac{\tell}{L}\right) \to 0 \quad \quad \text{as}\quad \quad \tell/ L\to 0
\end{split}
\end{align}
where we have used the fact that $\bbf$ is smooth to perform a Taylor expansion of the increment. Finally, the viscous term is bounded using integration-by-parts as follows:
\begin{align}
\| \nu\Delta\overline{\bu}''  \|_p &\leq \frac{\nu}{\oell^2} \int d \br \ |(\Delta G)_\ell(\br)|\| \delta \widetilde{\bu}(\br;\cdot)\|_{p} = \mathcal{O}\left( \frac{\nu}{\oell^2}\| \bu\|_{p} \right)\ \overset{\nu\to 0}{\longrightarrow}  \ 0\label{viscTerm}
\end{align}
and is thus negligible in the limit $\nu\to 0$ ($Re\to \infty$) in a fixed band $[\oell,\tell]$.  The bound \eqref{eq:sizeF} follows from collecting the above results.  

Note that the estimates for both the forcing and the viscous terms are asymptotically much smaller than the other bounds (i.e. for $Re\gg 1$ and $\tell/L\ll 1$).  It is also worth remarking that $\nu\Delta\overline{\bu}''$ is the only term in  $\bbF$ for which it is necessary to band-pass filter.  It we instead decomposed the velocity into a coarse and fine-grained field $\bu = \overline{\bu}+\bu'$ and identified small-scale motion as $\bu'$, the contribution of the corresponding viscous term cannot be rigorously neglected at high Reynolds number since derivatives of $\bu'$ are dissipation range quantities (i.e. there is an unregulated divergence $\Delta{\bu}'\to \infty$ as $\nu\to 0$).

\section{Direct Numerical Simulations} \label{appendix:DNS}

The DNS {of homogeneous, isotropic turbulence} were performed using a standard pseudo-spectral solver, based on the vorticity formulation of the Navier-Stokes equations.
Time stepping is performed by means of a memory-saving third-order Runge-Kutta method \cite{shu1988jcp}, and aliasing errors are controlled with a high-order Fourier smoothing \cite{hou2007jcp}.
Two simulations were run in cubic boxes of edge size $2\pi$, using $2048^3$ grid nodes for the real space grid: a longer DNS of just the fluid equations, with snapshots saved at regular intervals to compute field statistics, and a shorter DNS where trajectories of tracers of the coarse-grained fields were also integrated.

A quasi-stationary regime was maintained by the use of a large-scale Lundgren type of forcing \cite{Lundgren2003arb}, i.e. $\bbf$ is used in the NS equations \eqref{NSeqn} with ${\bbf} = \gamma {\bold{\hat\bu}}$, where ${\bold{\hat\bu}}$ is the band-passed velocity field restricted to a discrete band of small Fourier modes $k \in [2,4]$ and $\gamma= 1/2$ (DNS units).
For pseudo-spectral simulations, the energy spectrum $E(k)$ and the dissipation $\epsilon$ are easily accessible, and we follow \cite{Ishihara2007} for further computing standard statistical quantities:
\begin{equation}
    U = \sqrt{\frac{2E}{3}}, \hskip .5cm
    L = \frac{\pi}{2U^2} \int \frac{dk}{k} E(k), \hskip .5cm
    \eta = \left(\frac{\nu^3}{\epsilon}\right)^{1/4}, \hskip .5cm
    \lambda = \sqrt{\frac{15 \nu U^2}{\epsilon}}, \hskip .5cm
    Re_{\lambda} = \frac{U \lambda}{\nu}.
\end{equation}

The scaling exponents of the $p=2,4,6$ structure functions are found to be
$\zeta_2 = 2 \times (0.355 \pm 0.005)$,
$\zeta_4 = 4 \times (0.32 \pm 0.01)$,
and
$\zeta_6 = 6 \times (0.28 \pm 0.01)$.
They were obtained using logarithmic derivatives, see Fig. \ref{fig:scaling_exponents}.
A strict scaling regime is not present for the full $30 \eta$ to $L$ interval, systematic departures appearing towards the ends of the interval.
However there are distinct regions where the curves are flat, such that scaling exponents with well-defined values are observed.

The two DNS have a $Re_\lambda \approx 437$, and a low resolution of $k_M \eta \approx 1$ required if both a good scale separation $L / \eta \approx 470$ and good decorrelation over the computational box ($2\pi / L > 11$ for these runs) are desired.
The longer DNS was integrated over approximately $11$ integral times $L/U$ (or a total of $\approx 480 \tau_\eta$, where the Kolmogorov time scale is $\tau_\eta = \left(\nu/\epsilon\right)^{1/2}$), while the shorter particle run was integrated over approximately $17\tau_\eta$.

\begin{figure}
    \includegraphics{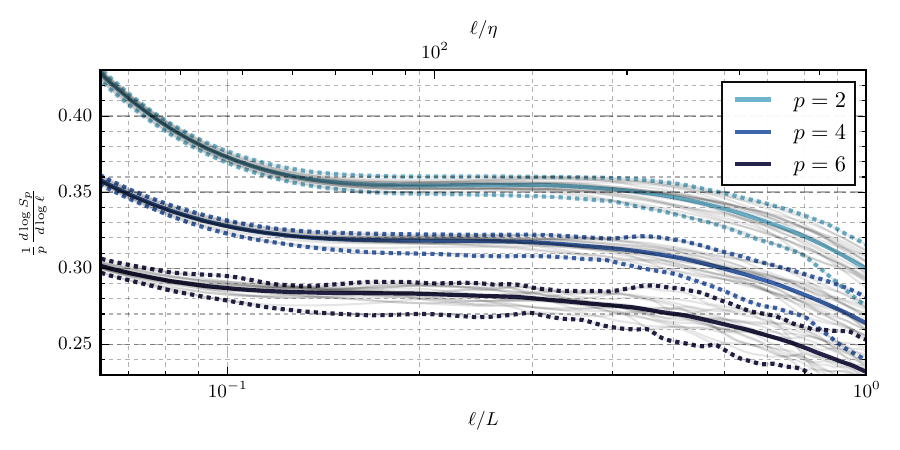}
    \caption{
        Logarithmic derivatives of the first three even-order structure functions, from a DNS at $Re_\lambda = 437$, shown for the range $30\eta < \ell < L$.
        Thin gray lines represent logarithmic derivatives obtained for individual snapshots of the DNS, the thick, continuous lines are obtained after averaging, and the thick dotted lines are the minimum and maximum over the different frames --- providing a way to ``bootstrap'' the error.
        Exponents are read directly from the plot, from the regions where the lines are flat --- and errors are only estimated from those same intervals.     }
    \label{fig:scaling_exponents}
\end{figure}

Trajectories of fluid tracers were computed using a second-order Adams-Bashforth method (see, e.g., \cite{atkinson1989book}) combined with cubic interpolation performed on four-point interpolation kernels.
Convergence tests with higher-order integration and interpolation schemes \cite{lalescu2010jcp} were performed, but are not reported here.
Several ``species'' of particles were used, {tracking fields coarse grained at different length scales}, so that the predictions of \eqref{eq:wscale} could be verified. Spatial filtering was performed using sharp spectral cutoffs for convenience.

\bibliography{references.bib}
\end{document}